# On the Four Types of Weight Functions for Spatial Contiguity Matrix

Yanguang Chen

（Department of Geography, College of Environmental Sciences, Peking University, Beijing 100871,

PRC. Email: chenyg@pku.edu.cn）

**Abstract:** This is a "spatial autocorrelation analysis" of spatial autocorrelation. I use the 1-dimension spatial autocorrelation function (ACF) and partial autocorrelation function (PACF) to analyze four kinds of weight function in common use for the 2-dimensional spatial autocorrelation model. The aim of this study is at how to select a proper weight function to construct a spatial contiguity matrix for spatial analysis. The scopes of application of different weight functions are defined in terms of the characters of their ACFs and PACFs.

## 1 Introduction

Spatial autocorrelation plays a very important role in spatial analysis of geographical systems (Haining, 2009). In fact, the analytical methods of spatial autocorrelation in geography fall into two classes--one is the 1-dimensional spatial autocorrelation based on the lag function without weight (Chen, 2008), and the other is the 2-dimension spatial autocorrelation based on the weight function without lag (Cliff and Ord, 1981). The latter, including Moran's *I* and Geary's *C*, is more familiar to geographers. One of the vital steps of spatial autocorrelation modelling is to construct a square spatial weights matrix (Getis, 2009). However, how to select weight function objectively is still a pending question remaining to be resolved. If we fail to choose a proper weight function, the result and effect of spatial analysis will not be satisfying and convincing, or even the calculation will be distorted (Chen, 2009). This letter will discuss four functions which can be used as weight function for 2-dimension spatial autocorrelation using the 1-dimension correlation function. The conclusions may be revealing for the students who plan to exercise spatial autocorrelation.

## 2 Weight functions

### 2.1 Spatial contiguity matrix

Before constructing a spatial weights matrix, we must make a spatial contiguity matrix by using weight function. For $n$ elements in a geographical system, a spatial contiguity matrix, $C$, can be expressed in the form

$$C = \begin{bmatrix} c_{11} & c_{12} & \cdots & c_{1n} \\ c_{21} & c_{22} & \cdots & c_{2n} \\ \vdots & \vdots & \ddots & \vdots \\ c_{n1} & c_{n2} & \cdots & c_{nn} \end{bmatrix}. \tag{1}$$

where $c_{ij}$ is a measurement used to compare and judge the degree of nearness or the contiguous relationships between region $i$ and region $j$. Thus a spatial weights matrix can be defined as

$$W = \frac{C}{C_0} = \begin{bmatrix} w_{11} & w_{12} & \cdots & w_{1n} \\ w_{21} & w_{22} & \cdots & w_{2n} \\ \vdots & \vdots & \ddots & \vdots \\ w_{n1} & w_{n2} & \ddots & w_{nn} \end{bmatrix} \text{ or } W_* = \frac{nC}{C_0} = n \begin{bmatrix} w_{11} & w_{12} & \cdots & w_{1n} \\ w_{21} & w_{22} & \cdots & w_{2n} \\ \vdots & \vdots & \ddots & \vdots \\ w_{n1} & w_{n2} & \ddots & w_{nn} \end{bmatrix}, \tag{2}$$

where

$$C_0 = \sum_{i=0}^{n}\sum_{j=0}^{n} c_{ij}, \quad \sum_{i=0}^{n}\sum_{j=0}^{n} w_{ij} = 1.$$

Obviously, $W$ is mathematically equivalent to $W_*$. In literature, $W_*$ used to act as spatial weights matrix. However, I argue that $W$ has an advantage of $W_*$ because we can build more regular model of spatial autocorrelation based on $W$.

### 2.2 Spatial weight function

Sometimes it appears as if there are two kinds of major 'forces' acting in natural world. The first force is one which tends to spread everything through space, while the second force, on the contrary, striving toward clustering and isolation. The first force leads to uniformity and equality on the whole, while the second force results in differentiability and inequality (El Naschie, 2000). The first force can be likened to the *action at a distance* on the earth surface, while the second force can be compared to *localization* in complex geographical system (Chen, 2008). The action at a distance reminds us the first law of geography of Tobler (1970, 2004), while localization suggests some force anti the first law of geography. The interaction of the two forces brings on four kinds of spatial action or correlation: (1) long-distance action or global correlation; (2) quasi-long-distance correlation action or quasi-global correlation; (3) short-distance action or quasi-local correlation; (4) proximal action or local correlation. The four types of action or correlation can be described with four different functions (Figure 1).

(1) Long-distance action or global correlation and inverse power function. The inverse power law reflects the action at a distance in geography, so the global correlation can be reflect with an inverse power function such as

$$c_{ij} = r_{ij}^{-b}. \tag{3}$$

where $b$ denotes the distance friction coefficient (generally, $b=1$). This kind weight function comes from the impedance function of the gravity model (Haggett *et al*, 1977). Cliff and Ord (1973, 1981) used this kind of function to construct spatial congruity matrix, which is sometimes called the Cliff-Ord weights.

(2) Quasi-long-distance correlation action or quasi-global correlation and negative exponential function. The quasi-global correlation can be described by the following exponential function

$$c_{ij} = \exp(-\frac{r_{ij}}{\bar{r}}). \tag{4}$$

where $r_{ij}$ refers to the distance between location $i$ and location $j$, and $\bar{r}$, the average distance of all the distances between any two locations. This function can be derived from the entropy-maximizing model of Wilson (1970).

(3) Short-distance action or quasi-local correlation and semi-step function. Short-distance action can be also called quasi-proximal action. Defining a critical distance $r_0$, the short-distance action can be describe by a semi-staircase function

$$c_{ij} = \begin{cases} 1, & \text{if } r_{ij} \leq r_0 \\ 0, & \text{if } r > r_0 \end{cases}, \tag{5}$$

which gives a binary matrix or 0-1 contiguity matrix. In fact, the semi-step function can be regarded as the special case of the negative exponential function. If $r_{ij} << \bar{r}$, then $c_{ij} \to 1$; while if $r_{ij} >> \bar{r}$ as given, then $c_{ij} \to 0$.

(4) Proximal action or local correlation and step function. Proximal action is an immediate

action or neighboring correlation. If region $i$ is next to region $j$, the value of spatial correlation is 1, otherwise, the value is 0. The local correlation can be described by the step function

$$c_{ij} = \begin{cases} 1, & \text{if region } i \text{ is on region } j \\ 0, & \text{others} \end{cases}. \tag{6}$$

This also gives a 0-1 contiguity matrix.

Now, a question arises--which weight functions should we choose? This depends on the type of spatial correlation or interaction in given study region. In order to clarify the criterion of selecting a weight function, we must know the similarities and differences between these weight functions.

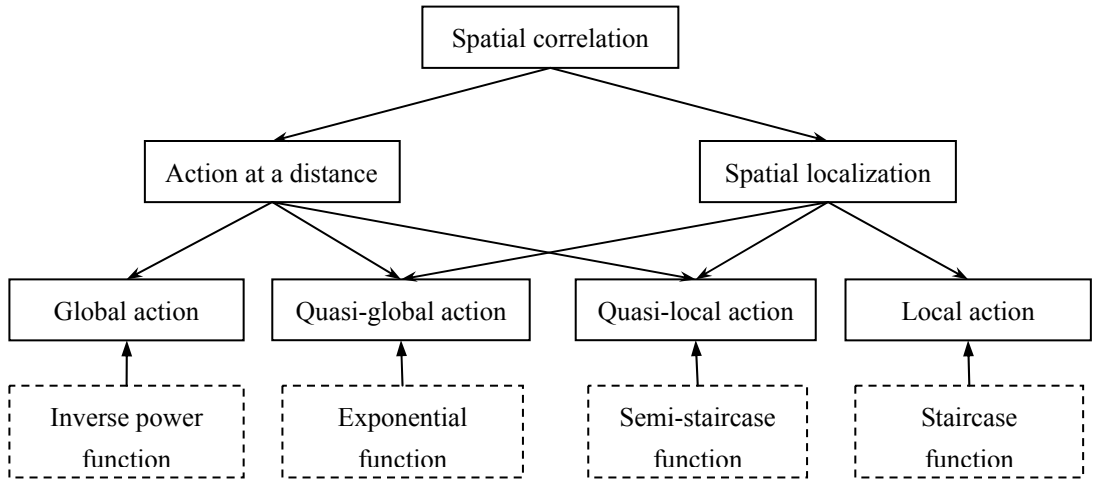

**Figure 1 The four types of spatial actions corresponding to four kinds of weight function**

## 3 Comparison and analysis

None is perfect in this world. Everything has its strong and weak points as a foot is oft-times too short and inch too long. We can employ the 1-dimension spatial autocorrelation analysis to reveal to advantages and disadvantages of different weight function for the 2-dimensional spatial autocorrelation analysis. The autocorrelation function (ACF) of a space series can be expressed as

$$R(k) = \frac{\sum_{i=k+1}^{N}[(c_i - \overline{c})(c_{i-k} - \overline{c})]}{\sum_{i=1}^{N}(c_i - \overline{c})^2}, \tag{7}$$

where $k$ refers to a displacement, or spatial lag, $R(k)$, to ACF corresponding to displacement $k$, $c_i$ represents a sample path of a spatial series with a length $N$, which can be generated by the above-given weight function, and $\bar{c}$, the average value of $c_i$, that is

$$\bar{c} = \frac{1}{N} \sum_{i=1}^{N} c_i. \qquad (8)$$

After calculating ACF, we can compute the partial autocorrelation function (PACF), which can be given by the well-known Yule-Walker equations. ACF implies both direct and indirect action, and PACF means only direct influence. You can find equation (7) and the Yule-Walker equation in any textbook on space/time series. It is no matter that if you have little idea of ACF and PACF and don't know how to work out them. If so, please utilize SPSS (Statistical Package for the Social Sciences) or other computer program used for statistical analysis. Actually, SPSS is to a mathematical beginner what a point-and-shoot camera is to a traveler knowing little about photography.

Now, we can generate four spatial series by using equations (3), (4), (5), and (6). Suppose the length of the sample path is $N=500$. First, let $b=1$ and $r=1, 2, \ldots, 500$, we can produce a simulative spatial series, $1/r$, for equation (3). Second, let $r=0, 1, 2, \ldots, 499$, and $\bar{r} = \sum_r r/499 = 249.5$, we can yield a spatial series, $\exp(-r/\bar{r})$, for equation (4). Third, let $r=0, 1, 2, \ldots, 499$, $r_0=4$, we can create a series such as 1, 1, 1, 1, 1, 0, 0,…0 for equation (5). Fourth, let $r=0, 1, 2, \ldots, 499$, it is easy for us to make a series in the form 1, 1, 0, 0,…0 for equation (6). Note that there are two "1" rather than one "1" at the beginning of the fourth series. In fact, when $k=0$, the correlation is with a region itself; when $k=1$, the correlation is with the neighbor. The calculations of ACFs and PACFs are shown in Figures 2, 3, 4, and 5, and the characters of these ACFs and PACFs are displayed in Table 1. For each histogram of ACF or PACF, the two transverse lines is called the "two-standard-error bands", according to which we can judge whether or not there is significant difference between ACF or PACF values and zero.

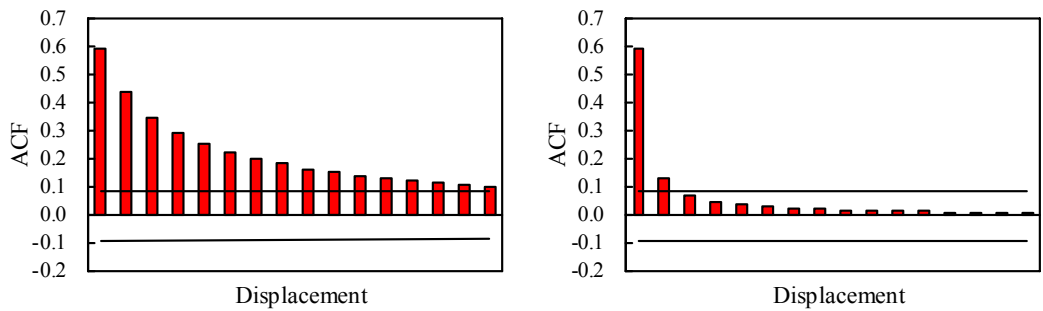

a. ACF histogram        b. PACF histogram

**Figure 2 The histograms of spatial ACF and PACF based on the inverse power-law distribution**

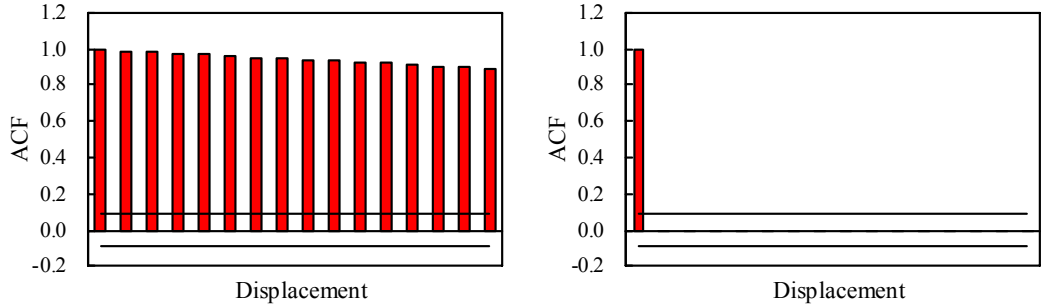

a. ACF histogram        b. PACF histogram

**Figure 3 The histograms of spatial ACF and PACF based on the negative exponential distribution**

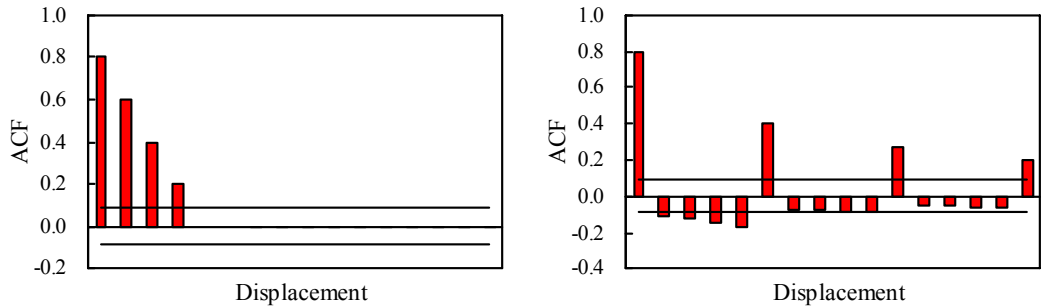

a. ACF histogram        b. PACF histogram

**Figure 4 The histograms of spatial ACF and PACF based on the semi-step-like distribution**

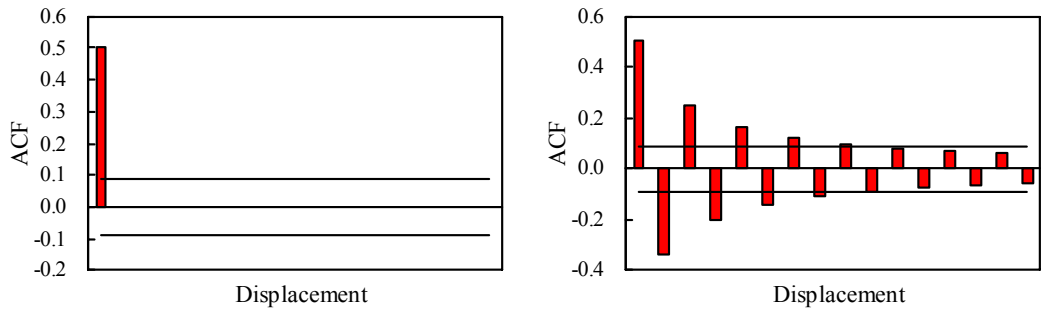

a. ACF histogram        b. PACF histogram

**Figure 5 The histograms of spatial ACF and PACF based on the step-like distribution**

Apparently, for the inverse power function, both ACF and PACF have trailing phenomena, the correlation functions die out little by little (Figure 2). In short, both ACF and PACF suggest long-distance action or global correlation. For the negative exponential function, ACT tails out, but PACF cuts off after displacement $k$=1(Figure 3). ACF suggests long-distance action or global correlation, while PACF suggest proximal action or local correlation. For the semi-step function, ACT cuts off at displacement $k$=5, but PACF periodically changes and dies away (Figure 4). ACF suggests short-distance action or quasi-local correlation, while PACF suggest quasi-distant effect and quasi-periodic change. PACF breaks out and through the "two-standard-error bands" every four "lags" until fading. For the step function, ACT cuts off after displacement *k*=1, but PACF displays a gradual two-sided damping and alternate effect (Figure 5). ACF suggests proximal action or local correlation, while PACF suggest distant effect and quasi-periodic change. PACF breaks out and through the "two-standard-error bands" every one "lag" until going out.

Table 1 Comparison of ACF and PACF between the four weight functions

| CF | Power function | Exponential function | Semi-step function | Step function |
|---|---|---|---|---|
| ACF | Tail off (global) | Tail off (global) | Cut off (quasi-local) | Cut off (local) |
| PACF | Tail off (global) | Cut off (local) | Tail off (quasi-period) | Tail off (period) |

## 4 Discussion and conclusion

The difference between inverse power function and the negative exponential function is clear. An exponential function suggests a characteristic scale ($\bar{r}$) and implies simplicity, while a power function suggests no characteristic scale and implies complexity (Chen, 2010). Batty and Kim (1992) made interesting discussion about exponential function and power function. Sometimes, exponential distributions indicate localization (Chen, 2008). The step function seems to be very simple, but it is actually complex because of oscillating and damping PACF. This suggests that step function is better for negative spatial autocorrelation than for positive spatial autocorrelation. The four weight functions can be compared with one another and the principal scopes of application of them are defined as below (Table 2).

Table 2 Comparison of spheres of application of the four weight functions

| Function | System type | Scale | Positive/Negative |
|---|---|---|---|
| Inverse power function | Complex systems | Large scale | Positive |
| Negative exponential function | Simple systems | Small scale | Positive |
| Semi-step function | Complex systems | Large scale | Positive or negative |
| Step function | Simple systems | Large scale | Negative |

The main conclusions of the letter can be drawn as follows. First, if the geographical phenomena follow a power law, you had better select the inverse power law as a weight function. The inverse power law is suitable for the large scale complex systems with global correlation. Second, if the geographical phenomena take on exponential distribution, you had better choose the negative exponential function as a weight function. The negative exponential function is fit for the medium-sized or small scale simple system with quasi-local correlation. Third, if the geographical phenomena are of localization, you had better select the semi-step or step function. The step function is appropriate for the region systems with local negative correlation. The semi-step function has a special importance in spatial scaling analysis, which will be discussed in a companion paper.